\documentclass[twocolumn,eqsecnum,aps]{revtex4}
\usepackage{epsfig}

\def\rhovec{\mbox{\boldmath $\rho$}}

\newcommand{\beq}{\begin{eqnarray}}
\newcommand{\eeq}{\end{eqnarray}}
\newcommand{\ket}{\rangle}
\newcommand{\bra}{\langle}

\begin{document}


\title {Four-body structure of $^7_{\Lambda}$Li
and $\Lambda N$ spin-dependent interaction}

\author{E.\ Hiyama}

\address{Department of Physics,
Nara Women's University, Nara 630-8506, Japan}

\author{Y. Yamamoto}

\address{Physics Section, Tsuru University, Tsuru, Yamanashi 402-8555, Japan}

\author{Th.A. Rijken}

\address{Institute of Mathematics, Astrophyiscs, and Particle Physics,\\
Radboud University, The Netherlands}

\author{T. Motoba}

\address{Laboratory of Physics, Osaka Electro-Comm.
University, Neyagawa 572-8530, Japan}

%


\begin{abstract}
Two spin-doublet states of
in $^7_{\Lambda}$Li are studied on the basis of the $\alpha +\Lambda +n+p$
four-body model. We employ the two-body interactions
which reproduce the observed properties of
any subsystems composed of $\alpha N$, $\alpha \Lambda$ and
$\alpha NN$, and $\alpha \Lambda N$.
Furthermore, the $\Lambda N$ interaction is adjusted so as to
reproduce the $0^+$-$1^+$ splitting of in $^4_{\Lambda}$H.
The calculated energy splittings of
$3/2^+$-$1/2^+$ and $7/2^+$-$5/2^+$ states in $^7_{\Lambda}$Li
are 0.69 MeV and 0.46 MeV, which are in good agreement with
the resent observed data.
The spin-dependent components of the $\Lambda N$ interaction
are discussed.
\end{abstract}

\maketitle

\section{Introduction}

It is a fundamental problem in  hypernuclear
physics to explore the features of underlying
interactions between hyperons ($Y$) and nucleons ($N$) through
analysis of many-body phenomena,
because $YN$ scattering data in free space are quite limited.
Then, quantitative analyses for light $\Lambda$ hypernuclei,
 where the features of $\Lambda N$ interactions
appear rather straightforwardly in observed level structures,
are of a special significance. In this connection, it is very important
that accurate measurements for $\gamma$-ray spectra have
been performed systematically \cite{Tamura00,Ajimura01,Akikawa02,
Ukai04,Ukai06}, which can be used to extract
the spin-dependent components of $\Lambda N$ interactions
through the detailed analyses of hypernuclear structures.
Although several shell-model calculations for light $\Lambda$
hypernuclei have been performed with the restricted
configuration of
$(0s)^4(0p)^n0s_{\Lambda}$ \cite{Gal71,Millener85,Richter91,Fetisov91},
their structures can be represented most excellently by the
cluster models. Today, it is possible to perform fully
microscopic calculations of three- and four-cluster systems
with sufficient numerical accuracy. Such sophisticated
calculations make it possible to study underlying
$\Lambda N$ interactions in comparison with the hypernuclear
data observed in the $\gamma$-ray experiments \cite{Tamura00,
Ukai06}. Because both  short- and long-range correlations
of the $\Lambda$ in nuclei are treated very accurately
in our approach, the characteristics  of the free-space $\Lambda N$
interactions can be studied very precisely.

The aim of this work is to analyze the ground $1/2^+$-$3/2^+$
and excited $5/2^+$-$7/2^+$ doublets in $^7_{\Lambda}$Li
keeping the consistency with the $5/2^+$-$3/2^+$
doublet in $^9_{\Lambda}$Be, and to determine the $\Lambda N$
spin-spin and spin-orbit interactions accurately
based on the experimental data for
$^9_{\Lambda}$Be, $^7_{\Lambda}$Li and $^4_{\Lambda}$H.
The splitting energies for the ground $1/2^+$-$3/2^+$ and
excited $5/2^+$-$7/2^+$ doublets in
$^7_{\Lambda}$Li \cite{Tamura00,Ukai06} are related
intimately to the spin-dependent potentials of the $\Lambda N$
interaction. Considered naively, the former splitting is
determined by the spin-spin interaction between the
$0s$-$\Lambda$ and the {\it deuteron} cluster, while the latter
is related to both spin-spin and spin-orbit interactions.
Thus, it is  critical whether or not this $5/2^+$-$7/2^+$
splitting in $^7_{\Lambda}$Li is reproduced consistently with those of
$^7_\Lambda$Li$(1/2^+$-$3/2^+)$ and
$^9_\Lambda$Be$(5/2^+$-$3/2^+)$.


Before starting a realistic calculation with the microscopic
four-body cluster model, we emphasize that the experimental
data of the $^7_{\Lambda}$Li and $^9_{\Lambda}$Be energy
levels are of a great value for  the $\Lambda$N
interaction study. First the low-lying state energies,
$^7_{\Lambda}$Li($1/2^+,3/2^+$), $^7_{\Lambda}$Li($5/2^+,7/2^+$),
$^9_{\Lambda}$Be($1/2^+$), and $^9_{\Lambda}$Be($3/2^+,5/2^+$),
are known recently with amazingly high resolution through the
$\gamma$-ray measurements\cite{Tamura00,Ajimura01,Akikawa02,Ukai04,
Ukai06}.
Secondly, the basic structure of these states are well understood
on the basis of the symmetry consideration without assuming
a specific form for the underlying $\Lambda$N interactions.

In order to verify level-energy consistency in the second point
mentioned above, let us make a preliminary calculation based
on the naive SU$_3$ wave functions.
In other words, we check whether the use of the low-lying state
energies known for $^7_{\Lambda}$Li ($1/2^+$, $3/2^+$, $5/2^+$) and
$^9_{\Lambda}$Be ($1/2^+$, $3/2^+$, $5/2^+$) leads to the right
position of the $^7_{\Lambda}$Li ($7/2^+$).
Based on the nuclear core wave functions
\begin{eqnarray}
\Phi_6(^6{\rm Li}:1_g^+, 3_1^+;T=0)=|[2](20)_{L=0,2}(S=1);J_c\ket \\
\Phi_8(^8{\rm Be};0_g^+,2_1^+,T=0)=|[4](40)_{L=0,2}(S=0);J_c\ket,
\end{eqnarray}
the hypernuclear states in $^7_{\Lambda}$Li and $^9_{\Lambda}$Be
can be assumed to have the configurations
with an $s$-state $\Lambda$ weak coupling:
 $[\Phi_N(L,S; J_c,T)\otimes\Lambda (s_{1/2})]_{J_H^+}$.
By using the Hamiltonian
${\widetilde H_A}=H_N+\epsilon_{\Lambda}(s_{1/2})+\sum{V_{N\Lambda}}$,
the hypernuclear level energies $\widetilde E_A(J_H)$ can be
expressed straightforwardly as

\begin{eqnarray}
\left[
\begin{array}{c}
  {\widetilde E_7(1/2^+)} \\
  {\widetilde E_7(3/2^+)} \\
  {\widetilde E_7(5/2^+)} \\
  {\widetilde E_9(3/2^+)} \\
  {\widetilde E_9(5/2^+)} \\
\end{array}
\right]
&=
\left[
\begin{array}{c}
  E_6(1_g^+)\\
  E_6(1_g^+) \\
  E_6(3^+) \\
  E_8(2^+) \\
  E_8(2^+) \\
  \end{array}
\right]
+\epsilon_{\Lambda}  \nonumber \\
&+
\left[
\begin{array}{ccccc}
\frac{5}{18}&\frac{19}{18}&\frac{11}{18}&\frac{1}{18}&\frac{-16\sqrt{2}}{27}\\
\frac{10}{9}&\frac{2}{9} &\frac{4}{9}&\frac{2}{9}&\frac{8\sqrt{2}}{27} \\
\frac{1}{4} & \frac{7}{4} &    0  &  0  &    0           \\
\frac{25}{24}& \frac{13}{8} &\frac{3}{4} &\frac{7}{12}&\frac{\sqrt{2}}{2}\\
\frac{25}{12}&\frac{7}{12}& \frac{7}{6}& \frac{1}{6}&\frac{-\sqrt{2}}{3}\\
  \end{array}
\right]
\left[
\begin{array}{c}
 v_a  \\ v_b \\ v_c \\ v_d \\ v_e \\
\end{array}
\right]
\end{eqnarray}
\noindent
where $v_a$, $v_b$, $v_c$, $v_d$, and $v_e$ stand for the
$N\Lambda$ interaction matrix elements \\
$\bra p_{3/2}s_{1/2}^{\Lambda}|V_{N\Lambda}|
p_{3/2}s_{1/2}^{\Lambda}\ket_{2^-}$,
$\bra p_{3/2}s_{1/2}^{\Lambda}|V_{N\Lambda}|
p_{3/2}s_{1/2}^{\Lambda}\ket_{1^-}$,\\
$\bra p_{1/2}s_{1/2}^{\Lambda}|V_{N\Lambda}|
p_{1/2}s_{1/2}^{\Lambda}\ket_{1^-}$,
$\bra p_{1/2}s_{1/2}^{\Lambda}|V_{N\Lambda}|
p_{1/2}s_{1/2}^{\Lambda}\ket_{0^-}$,\\
 and
 $\bra p_{3/2}s_{1/2}^{\Lambda}|V_{N\Lambda}|
p_{1/2}s_{1/2}^{\Lambda}\ket_{1^-}$,
respectively. If we input the experimental energies for
$E_6$, $E_8$, $\widetilde E_7$, and $\widetilde E_9$, 
a set of the interaction matrix elements \{$v'$s\} are obtained.
Then, one finds that the use of the solution \{$v'$s\} leads to
the theoretical result of 2.53 MeV for $\widetilde E_7(7/2^+)$
and also that this result is quite consistent with the
experimental value $2.521\pm 0.04$ MeV\cite{Ukai06}.
Thus these energy levels of $^7_{\Lambda}$Li and $^9_{\Lambda}$Be
are surely based on the similar maximum spatial symmetry, and
therefore a detailed realistic calculation with the microscopic
cluster model should have an important value.

In the past, two types of cluster model calculations
have been performed on the basis of the
$\alpha+d+\Lambda$ and $^5_{\Lambda}$He$+n+p$ configurations.
 Using the $\alpha +d+\Lambda$ cluster
model\cite{Motoba83,Wang87,Yamamoto91}, it has been
discussed that the ground $1/2^+_1$-$3/2^+_1$ doublet in
$^7_{\Lambda}$Li is an important candidate
to extract the $\Lambda N$ spin-spin interaction.
In Ref. \cite{Hiyama96}, we proposed the
$^5_{\Lambda}{\rm He}+n+p$ model, where the full $n-p$
correlation is taken into account without the frozen-deuteron
cluster approximation. The splitting energies of the ground
and excited doublets in $^7_{\Lambda}$Li were calculated to
be 0.87 MeV and 0.81 MeV, respectively.

As for the spin-orbit splittings in $^9_{\Lambda}$Be
and $^{13}_{\Lambda}$C, the three-body ($2\alpha +\Lambda$)
model and the four-body ($3\alpha +\Lambda$) model have
been applied, respectively \cite{Hiyama00}.
The calculated splitting energies of
$^9_{\Lambda}$Be$(5/2^+_1$-$3/2^+_1)$ were 80-200 keV
when we adopted the symmetric (SLS) and anti-symmetric
(ALS) spin-orbit interactions derived from the Nijmegen
OBE models. These theoretical values are considerably
larger than the experimental value $43 \pm 5$ keV \cite{Akikawa02}.
At the same time\cite{Hiyama00}, however, we tried to
enlarge the ALS potential to be 85 \% of the SLS as
inspired by the a quark-model $\Lambda N$ interaction,
we predicted the smaller splitting of 35-40 keV before
the experiment. The similar discussion was given also
to explain the small spin-orbit splitting of the $\Lambda$
$p$-state observed in $^{13}_{\Lambda}$C\cite{Ajimura01,Akikawa02}.

In this work, we extend these cluster models to the
four-body treatment of $\alpha+n+p+\Lambda$ so as to take
account of the full correlations among all the constituent
particles. Such an extended calculation has been tried
once in Ref. \cite{Hiyama98}. Here we focus our
attention especially to two spin-doublets in
$^7_{\Lambda}$Li, keeping the consistency with that
in $^9_{\Lambda}$Be. In order to extract the dynamical
information on the underlying $\Lambda N$ spin-spin
and spin-orbit interactions, two-body interactions among
constituent particles ($\alpha$, $n$, $p$, $\Lambda$)
are chosen so as to reproduce accurately the observed
properties of all the subsystems composed of $\alpha N$,
$\alpha \Lambda$ and $\alpha NN$, and $\alpha \Lambda N$.

\section{Four-Body Cluster Model and Method}

In this work, the hypernucleus $^7_{\Lambda}$Li
is considered to be composed of $\alpha$ cluster,
$\Lambda$ particle and two nucleons ($N$).
The core nucleus $\alpha$ is
considered to be an inert core and to have
the $(0s)^4$ configuration,
$\Psi(\alpha)$.
The Pauli principle between the valence nucleon
and the core nucleons is taken into
account by the orthogonality condition model
(OCM) \cite{Saito69}, as the valence nucleon's wavefunction should
be orthogonal to that the core nucleon.

Nine set of the Jacobian coordinates of the four-body
system of $^7_{\Lambda}$Li are illustrated in Fig.1
in which we further take into account the antisymmetrization
between the two nucleons.
\begin{figure}[htb]
\begin{center}
\epsfig{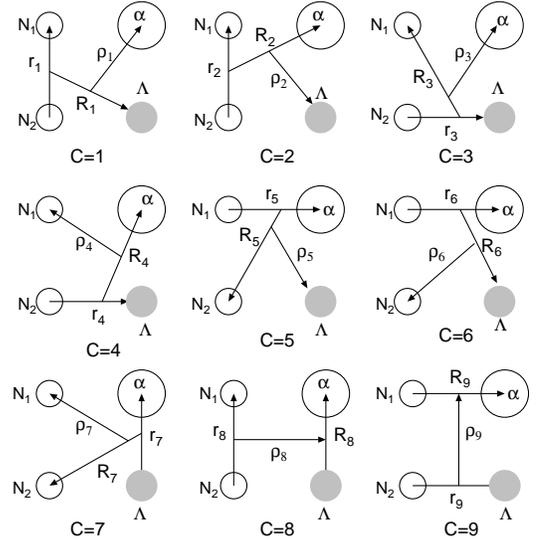}
\end{center}
\caption{Jacobian coordinates for all
the rearrangement channels ($c=1 \sim 9$)
of the $\alpha +\Lambda +n+p$ four-body model.
Two nucleons are to be antisymmetrized.}
\label{fig:level1}
\end{figure}
The Schr\"{o}dinger equation
are given by
\begin{eqnarray}
 && ( H - E ) \, \Psi_{JM}(^7_{\Lambda}{\rm Li})  = 0 \ , \\
 &&  H=T+\sum_{a,b}V_{a b}
      +V_{\rm Pauli} \ ,
\end{eqnarray}
where $T$ is the kinetic-energy operator and
$V_{\rm ab}$ is the interaction
between the constituent particles $a$ and $b$.
The Pauli principle between
the  $\alpha$ particle and two nucleons
is taken into account by the
Pauli projection operator
$V_{\rm Pauli}$ which is explained
in the next section as well as  $V_{a b}$.
The total wavefunction is described
as a sum of amplitudes of the rearrangement channels
Fig. 1 in the LS coupling scheme:
\begin{eqnarray}
      \Psi_{JM}\!\!&(&\!\!^A_{\Lambda}{\rm Li})
       =  \sum_{c=1}^{9}
      \sum_{n,N,\nu}  \sum_{l,L,\lambda}
       \sum_{S,\Sigma,I,K}
       C^{(c)}_{nlNL\nu\lambda S\Sigma IK} \nonumber  \\
      &  \times & {\cal A}_N
      \Big[
       \Phi(\alpha) \big[\chi_s(\Lambda)
     \big[ \chi_{\frac{1}{2}}(N_1)
       \chi_{\frac{1}{2}}(N_2)
       \big]_S \big]_\Sigma  \nonumber \\
  & \times  &    \big[ \big[ \phi^{(c)}_{nl}({\bf r}_c)
         \psi^{(c)}_{NL}({\bf R}_c)\big]_I
        \xi^{(c)}_{\nu\lambda} (\mbox{\boldmath $\rho$}_c)
        \big]_{K}  \Big]_{JM}  \;  .
\end{eqnarray}

Here the operator  ${\cal A}_N$
stands for antisymmetrization between the two
nucleons.
$\chi_{\frac{1}{2}}(N_i)$
and $\chi_{\frac{1}{2}}(\Lambda)$ are the spin functions of
the
$i$-th nucleon  and $\Lambda$ particle.
Following the Gaussian Expansion Method (GEM)
\cite{Kami88,Kame89,Hiyama03},
we take the functional form of
$\phi_{nlm}({\bf r})$,
$\psi_{NLM}({\bf R})$ and
$\xi^{(c)}_{\nu\lambda\mu} (\mbox{\boldmath $\rho$}_c)$ as
\begin{eqnarray}
      \phi_{nlm}({\bf r})
      &=&
      r^l \, e^{-(r/r_n)^2}
       Y_{lm}({\widehat {\bf r}})  \;  ,
 \nonumber \\
      \psi_{NLM}({\bf R})
      &=&
       R^L \, e^{-(R/R_N)^2}
       Y_{LM}({\widehat {\bf R}})  \;  ,
 \nonumber \\
      \xi_{\nu\lambda\mu}(\mbox{\boldmath $\rho$})
      &=&
       \rho^\lambda \, e^{-(\rho/\rho_\nu)^2}
       Y_{\lambda\mu}({\widehat {\rhovec}})  \; ,
\end{eqnarray}
where the Gaussian range parameters are chosen to lie
in geometrical progressions:
\begin{eqnarray}
      r_n
      &=&
      r_1 a^{n-1} \qquad \enspace
      (n=1 - n_{\rm max}) \; ,
\nonumber\\
      R_N
      &=&
      R_1 A^{N-1} \quad
     (N \! =1 - N_{\rm max}) \; ,  
\nonumber\\
      \rho_\nu
      &=&
      \rho_1 \alpha^{\nu-1} \qquad
     (\nu \! =1 - \nu_{\rm max}) \; .  
\end{eqnarray}
These basis functions have been verified to be suited for
describing both the short-range correlations
and the long-range tail behaviors of few-body systems
\cite{Kami88,Kame89,Hiyama03}.
  The eigenenergy $E$  in Eq.(2.1) and the
coefficients $C$ in Eq.(2.3) are to be determined by
the Rayleigh-Ritz variational method.

The angular-momentum space of the wavefunction
with $l, L, \lambda \leq 2$ was found to be enough
for getting satisfactorily convergence of the binding energies
of the states studied below
(note that no truncation is taken of the {\it interactions}
in the  angular-momentum space).  As for the numbers
 of the Gaussian basis,
$n_{\rm max}, N_{\rm max}$ and
$ \nu_{\rm max}$, $4 -10$ are enough.

\section{Interactions}

\subsection{$\alpha N$ interaction}

For the interaction $V_{N \alpha}$ between $\alpha$
and a valence nucleon, we employ the effective potential
proposed in Ref.\cite{Kanada79}, which is designed so as to reproduce
well the low-lying states and low-energy scattering phase
shifts of the $\alpha n$ system.

The Pauli principle between nucleons belonging to
$\alpha$ and valence nucleons is taken into account
by the orthogonality condition model (OCM) \cite{Saito69}.
The OCM projection operator $V_{\rm Pauli}$ is
represented by
\begin{equation}
V_{\rm Pauli}=\lim_{\lambda\to\infty} \ \lambda \
|\phi_{0s}({\bf r}_{N \alpha})
\rangle \langle \phi({\bf r}'_{N \alpha})|
\end {equation}
which excludes the amplitude of the
Pauli forbidden state $\phi_{0s}({\bf r})$ from
the four-body total wavefunction \cite{Kukulin84}.
The Gaussian range parameter $b$
of the single-particle $0s$ orbit in the
$\alpha$ particle is taken to be $b=1.358$ fm so as to
reproduce the size of the $\alpha$ particle.
In the actual calculation, the strength $\lambda$
for $V_{\rm Pauli}$ is taken to be $10^5$ MeV,
which is large enough to push
the unphysical forbidden states into the
very high energy region while keeping
the physical states unchanged.
Usefulness of this Pauli operator
method of OCM has been verified in many
cluster-model calculations.

\subsection{$NN$ interaction}

In order to study the fine structure of
our $\alpha +n+p+\Lambda$
system ($^7_{\Lambda}$Li),
it is necessary to use an $NN$
interaction
which reproduces accurately the energy spectrum
of the $\alpha +n+p$
subsystem ($^6$Li).
Such an $NN$ interaction is given here as
follows:
We start from the AV'8 \cite{Pudliner97} potential,
$V_{NN}$ ,
derived from the AV18 \cite{Wiringa95}
by neglecting the $(L \cdot S)^2)$
term.
In our model,
this potential gives the calculated
values of $-3.38$ MeV and
$-0.98$ MeV for the $1^+$ and $3^+$
states of $^6$Li, respectively,
being of less binding compared
to the experimental data.
Next, we adjust the central
and tensor parts of $V_{NN}$
together with the slight modification
of $V_{N \alpha}$ so that the
experimental energies of $^6$Li ($1^+,3^+)$
and deuteron are reproduced.

\subsection{$\alpha \Lambda$ interaction}

The interaction between the $\Lambda$ particle
and $\alpha$ cluster is derived
by folding the $\Lambda N$ G-matrix interaction
with a three-range Gaussian form
into the density of the
$\alpha$ cluster in the same manner as our
previous work in Ref.
\cite{Hiyama97}.
In the present work, we employ the G-matrix interaction
for Nijmegen model F(NF)~\cite{NDF}, the parameters of
which are also listed in Ref. \cite{Hiyama97}.
Even if the versions for the other Nijmegen models are used,
the obtained results are almost the same as the present one.
This is because our $\Lambda N$ folding interaction is
adjusted so as to reproduce the experimental value of
$B_\Lambda(^5_\Lambda$He).


\subsection{$\Lambda N$ interaction}

For $\Lambda N$ interactions, meson-theoretical models
have been proposed on the basis of the SU(3) symmetry
of meson-baryon coupling constants.
In principle, these realistic interactions can be used
directly in our four-body model of $^7_\Lambda$Li.
However, the purpose of this work is to extract the
information on the spin-dependent parts of the $\Lambda N$
interaction as quantitatively as possible using the
measured splitting energies of spin-doublet states.
We employ effective $\Lambda N$ single-channel
interactions simulating the basic features of the
Nijmegen meson-theoretical models
NSC97f~\cite{NSC97}, in which some potential parameters
are adjusted phenomenologically so as to reproduce
the experimental data.

Our $\Lambda N$ interactions, composed of
central, SLS and ALS parts
are represented as
\begin{eqnarray}
V^{({\rm C})}_{\Lambda N}(r) = \sum _{\alpha}
\sum^{3}_{i=1} v_i^{(\alpha)}
\ \exp{(-(r/\beta_i)^2)}  \quad ,
\end{eqnarray}
\begin{eqnarray}
V^{\rm LS}_{\Lambda N} = V^{SLS} {\bf L} {\bf S_+}
+ V^{ALS} {\bf L} {\bf S_-}
\end{eqnarray}
\noindent
with ${\bf S_{\pm}=s_{\Lambda} \pm s_N}$.
Here, the central potential, $V^{\rm C}_{\Lambda N}$,
with three-range Gaussian forms
are given separately for spin-parity states of
$\alpha$= $^3E$ (triplet even),
$^1E$ (singlet even), $^3O$ (triplet odd), $^1O$ (singlet odd).
The even- and odd-state spin-spin interaction are defined by
$(V_{\Lambda N}^{(^3E)}-V_{\Lambda N}^{(^1E)})/4$
and $(V_{\Lambda N}^{(^3O)}-V_{\Lambda N}^{(^1O)})/4$,
respectively.

The potential parameters in the central parts are chosen
so as to simulate $\Lambda N$ scattering
phase shifts calculated by NSC97f.
The determined parameters are given in Table I.
It should be noted here that
the $\Lambda N$-$\Sigma N$ coupling interactions
are included explicitly in NSC97f,
and their contributions in many-body systems are
different from those in free space.
This means that our obtained phase-shift equivalent potential
should be modified appropriately in applications
to hypernuclear system:  We adjust the second-range strengths
$v_2^{(^3E)}$ and $v_2^{(^1E)}$
so that calculated energies of
$0^+$-$1^+$ doublet state in our $NNN\Lambda$
four-body calculation reproduce the experimental values
obtained by those of $^4_\Lambda$H.
In Table I, the adjusted values of $v_2^{(^1E)}$ and
$v_2^{(^3E)}$ are shown in parentheses.
On the other hand,
there was no clear experimental data to determine
quantitatively the odd-state parts, which leads to
remarkable differences among theoretical interaction models.
Our present analysis for the splitting energies of $^7_\Lambda$Li
gives some constraint on the odd-state spin-spin part.
The second-range values of $v_2^{(^1O)}$ and $v_2^{(^3O)}$ in
parentheses are determined on the basis of the $^7_\Lambda$Li data,
as shown later.

\begin{table}[htbp]
  \caption{Parameters of the
  $\Lambda N$ interaction defined in Eq.(3.2).
  Range parameters are in fm and strength are in MeV.
  The numbers in parentheses are improved even state strength so as to
  reproduce observed spin doublet state in $^4_{\Lambda}$H
  and odd state
  strength so as to reproduce observed ground  doublet state
  in $^7_{\Lambda}$Li
  .}
  \label{tab:1}
\begin{center}
  \begin{tabular}{cccccc}
\hline
\hline
&$i$     & 1  &2 &3 \\
&$\beta_i$    &1.60   &0.80   &0.35  \\
\hline
       &$v_i{(^1E)}$ &--7.87  &--342.5 (--357.4) &  6132. \\
       &$v_i{(^3E)}$ &--7.89  &--242.4 (--217.3) &  3139. \\
       &$v_i{(^1O)}$ &--1.30  &  213.7 (\enskip 513.7) &  8119. \\
       &$v_i{(^3O)}$ &--3.38  &  122.9 (\ 22.9)  &  5952. \\
 \hline
  \end{tabular}
\end{center}
\end{table}

The SLS and ALS interactions here are chosen so as to
reproduce the $^9_\Lambda$Be data.
In Ref.\cite{Hiyama00}, the various sets were derived from
the Nijmegen models.
However, the $5/2^+$-$3/2^+$ splitting energies
obtained from these sets are considerably larger than
the experimental value.
Now, our SLS and ALS interactions are derived as follows:
Firstly, the SLS part derived from NSC97f with the G-matrix
procedure is represented in the two-range form
$V^{SLS}=\sum^{2}_{i=1} v^{(+)}_{i}\ e^{-(r/\gamma_i)}$
The values of the parameters are
$v_1^{(+)}= -110.6$ MeV and $v_2^{(+)}=-1157.$ MeV for
$\gamma_1=0.70$ fm and $\gamma_2=0.40$ fm, respectively
as given in Ref.\cite{Hiyama00}.
Next, assuming $V^{ALS}= -\alpha V^{SLS}$,
the parameter $\alpha$ is chosen so as to reproduce
the measured $5/2^+$-$3/2^+$ splitting energy with the
$2\alpha + \Lambda$ cluster model developed
in Ref.\cite{Hiyama00}.
Our obtained value is $\alpha=0.83$.
Using these $V^{SLS}$ and $V^{ALS}$,
we also calculated the energy splitting of
$1/2^-$-$3/2^-$ doublets in $^{13}_{\Lambda}$C based on
$3\alpha +\Lambda$ four-body model to be
0.2 MeV which is consisten with the observed data
within the error \cite{Ajimura01}.
This ALS interaction is fairly stronger than that
derived from NSC97f.
As discussed later, however, the similar result can be
obtained by weakening the SLS part without changing
the ALS part, because only the sum of SLS and ALS is
fixed by the $^9_\Lambda$Be data.
There is a famous quark-model prediction \cite{Morimatsu}
that
the ALS is so strong as to substantially
cancel the LS one.
It should be noted that this prediction is not necessarily
proved by our present analysis.

\section{Results}

\begin{figure}[htb]
\begin{center}
\epsfig{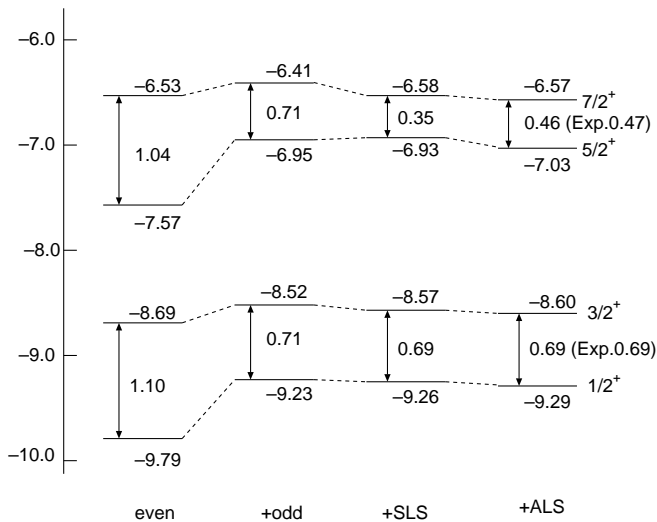}
\end{center}
\caption{Calculated energy levels of
$^7_{\Lambda}$Li on the basis of
$\alpha +\Lambda +n+p$ model.
The energies are measured from the $\alpha +\Lambda+n+p$ threshold.
The observed energy splittings of $3/2^+$-$1/2^+$ and
$7/2^+$-$5/2^+$ are 0.69 MeV and 0.47 MeV, respectively.}
\label{fig:level2}
\end{figure}

In Fig.2, we illustrate our result for the
$1/2^+$-$3/2^+$ and $5/2^+$-$7/2^+$ doublet states
of $^7_{\Lambda}$Li .
The energies of the $1^+$-$3^+$ doublet state of $^6$Li nucleus
calculated in the framework of the $\alpha\!+\!n\!+\!p$ three-body
model are $-3.7$ MeV and $-1.6$ MeV, being  measured from the
$\alpha\!+\!n\!+\!p$ three-body threshold.
As shown in the left side of the figure,
the calculated splitting energies for both doublets
are about 1 MeV very similar to that for $0^+$-$1^+$
doublet state of $^4_{\Lambda}$H ($^4_{\Lambda}$He),
when only the even-state central interaction is used.
Namely, the even-state spin-spin interaction turns out
to contribute similarly to the $0^+$-$1^+$ splitting
energy of $^4_{\Lambda}$H ($^4_{\Lambda}$He) and the
$1/2^+$-$3/2^+$ and $5/2^+$-$7/2^+$ ones of $^7_{\Lambda}$Li.

Next, let us switch on the odd-state central interaction.
When only the even-state interaction is used,
the obtained value of the ground-state energy
is $-9.79$ MeV.
When we use the $^1O$ and $^3O$ interactions derived
from NSC97f, the ground $1/2+$ state is obtained at $-9.23$
MeV. This energy changes only slightly ($-0.06$ MeV)
with the inclusion of SLS and ALS,
because the spin-orbit interactions have essentially
no effect on the $1/2+$ state due to its $L=0$
structure. The final value $-9.29$ MeV
means that the experimental $\Lambda$ binding energy
(5.58 MeV) is  reproduced well, because the calculated
energy of the $\alpha+p+n$ subunit is $-3.7$ MeV
in our model.
Found in the figure, this fact is owing to the
peculiar role of our odd-state interaction.
Therefore, the repulsive contribution
from the spin-independent part
$(3 V_{\Lambda N}^{(^3O)}+V_{\Lambda N}^{(^1O)})/4$
turns out to be decisive to reproduce the experimental value.
The repulsive nature of this part is an important property
of NSC97f.
On the other hand, the corresponding part in the recent model
ESC04~\cite{ESC04} is attractive in contrast to NSC97f.
The important role of the repulsive odd-state interaction
in our analysis does not necessarily mean that
the odd-state part in NSC97f is more realistic than
the one in ESC04, as discussed later.

As for the $1/2^+$-$3/2^+$ splitting, the
addition of the NSC97f odd-state central interaction
leads to 0.97 MeV. This splitting is too large in view of
the experimental value
(0.69 MeV), because the contribution of the SLS/ALS
interactions to this $1/2^+$-$3/2^+$ splitting is quite small.
We add here the attractive (repulsive) correction on the
$^3O$ ($^1O$) state interaction, which works efficiently
in high-spin (low-spin) partners of doublets, by making
the spin-spin interaction more attractive:
We introduce the attractive spin-spin interaction in
second range (0.8 fm),
$\Delta v_{ss} =-100.0$ MeV.
The modified values of $v_2^{(^1O)}$ and $v_2^{(^3O)}$
are shown in parentheses of Table.I, which leads to
the calculated values of 0.71 MeV and
0.54 MeV for the lower
and higher doublets, respectively (cf. Fig.2),
as seen in the figure.

Now, we come to the important stage of  looking  at the
roles of the SLS and ALS interactions for splitting energies.
It should be noted here that these interactions work
differently for the two doublet states of $^7_\Lambda$Li:
The contributions to the ground-state $1/2^+$-$3/2^+$ doublet
are very small, where the $pn$ pair part outside the
$\alpha$ core is dominated by the $L=0$ component spatially.
On the other hand, in the case of the excited $5/2^+$-$7/2^+$
doublet composed of the $L=2$ $pn$ pair,
the SLS and ALS interactions play important roles:
As seen in Fig.2, the SLS works  attractively
(slightly repulsively) for the $7/2^+$ ($5/2^+$) state,
because the $7/2^+$ state is dominated by the spin-triplet
configuration of the $L=2$ $pn$ pair and the $s$-state $\Lambda$.
On the other hand, the ALS works efficiently in the $5/2^+$ state
which has both configurations of spin-triplet and spin-singlet.
The ALS which acts between $S=0$ and $S=1$
$\Lambda N$ two-body states has essentially no
effect on the $7/2^+$ state.

Thus, owing to the combined effects of the SLS and ALS,
our final result reproduces nicely the observed
energies of the spin-doublet states in $^7_\Lambda$Li.
Recently, Millner calculate $3/2^+$-$1/2^+$ and $7/2^+$-$5/2^+$
spin-doublets states using a shell model 
with the phenomenological interaction matrix element \cite{Ukai06,Millener05}.
His calculation is also in good agreement with the
recent data.
It should be noted here that the strength of the ALS part
is not necessarily determined by our present analysis.
The above result is obtained by making the ALS part
stronger than that given by NSC97f so as to reproduce
the $^9_\Lambda$Be data. The very similar result,
however, can be obtained by weakening the SLS part
without changing the ALS part.


Before going to summary, we would like to comment on
the role of the $\Lambda N$-$\Sigma N$ coupling.
%
Our basic assumption in this work is that the
$\Lambda N$-$\Sigma N$ coupling interaction can be
renormalized into the $\Lambda N$-$\Lambda N$ interaction
effectively. In this spirit, the even-state parts of
our $\Lambda N$-$\Lambda N$ interaction were adjusted
so as to reproduce the $0^+$ and $1^+$ of $^4_\Lambda$H.
It is reasonable, however, to consider that
the $\Lambda N$-$\Sigma N$ coupling works more
repulsively in $^7_\Lambda$Li.
It is likely that the role of the odd-state repulsion
in our treatment is a substitute for this effect.
This is the reason why the attractive odd-state interaction
in ESC04 models cannot be denied.
As shown in Fig.2, the energy of $5/2^+$ state
is located above by about 0.2 MeV in comparison with the
observed energy of $5/2^+$ state.
This problem may be solved by taking into account
the repulsive effect of the
$\Lambda N$-$\Sigma N$ coupling
instead of the
odd-state repusion,
because the SLS/ALS interaction
works more efficiently under the attractive
odd-state interactions.
Some authors~\cite{Gibson72,Akaishi00} pointed out
the extra contribution to the $^4_\Lambda$H($0^+$-$1^+$)
splitting energy from the three-body correlated
$\Lambda N$-$\Sigma N$ mixing.
The present authors also obtained the value of
about 0.3 MeV for the three-body contribution
of $\Lambda N -\Sigma N$ coupling in the
$0^+$-$1^+$ splitting energy in $^4_{\Lambda}$H \cite{Hiyama01}.
In the shell model calculation \cite{Millener05},
Millener calculated the spin-doublets states in
$^7_{\Lambda}$Li including $\Lambda N -\Sigma N$ coupling
and he concluded that this contribution was
small in these splitting energies.
On the other hand, Fetisov pointed out  that
the large effect of $\Lambda N-\Sigma N$ coupling was
found in both of $^4_{\Lambda}$H and $^7_{\Lambda}$Li \cite{Fetisov99}.
It is an open problem to study $\Lambda N$-$\Sigma N$ coupling
effects consistently for $^4_\Lambda$H and $^7_\Lambda$Li.

\bigskip
In summary,
we discussed the two spin-doublets of $3/2^+$-$1/2^+$ and
$7/2^+$-$5/2^+$ in $^7_{\Lambda}$Li based on
$\alpha +\Lambda +n+p$ four-body model. Here,
it is important that all the two-body interactions are
chosen so as to reproduce both the binding energy of any
subsystem composed of two- and three-constituent particles.
Our $\Lambda N$ interactions, simulating $\Lambda N$
scattering phase shifts calculated by NSC97f,
are adjusted so as to reproduce the observed data
of spin-doublet states.
It is found that the even-state $\Lambda N$ interaction
leads to the similar values of the splitting energies of
the $0^+$-$1^+$ doublet in $^4_\Lambda$H ($^4_\Lambda$He)
and the $1/2^+$-$3/2^+$ and $5/2^+$-$7/2^+$
doublets in $^7_\Lambda$Li. Then,
the odd-state interactions play important roles
to reproduce the difference between the two doublet states
in $^7_\Lambda$Li.
With use of the SLS and ALS interactions adjusted so as to
reproduce the $5/2^+$-$3/2^+$ splitting in $^9_{\Lambda}$Be,
the two doublet states in $^7_\Lambda$Li can be reproduced
exactly by tuning the odd-state spin-spin interaction.

The basic assumption in our present approach is that
the $\Lambda N$-$\Sigma N$ coupling interactions are
renormalized reasonably into our $\Lambda N$ interactions.
The validity of this assumption will be investigated
in our future studies.
The coupled four-body calculation of $\alpha+\Lambda+N+N$ and
$\alpha +\Sigma + N+N$ is in progress.

Authors would like to thank Prof. H. Tamura, Prof. M. Kamimura, Prof.
T. Yamada and Prof. M. Tanifuji for helpful
discussions. This work has been supported in part by
Grant-in-Aid for Scientific Research  (17740145) of
Ministry of Education, Science and Culture of Japan and by the
 Nara Women's University Intramural Grant for Project Research
 (E. H.). One of the authors (Th.A. R.) is grateful to
Osaka Electro-Communication University for the hospitality for
three months, and another one (Y.Y.) thanks Radboud University 
Nijmegen for its hospitality.

\end{document}